# First-principles study of the stability of bimetallic PdPt nanoparticles under a finite temperature

Takayoshi Ishimoto[1,2] and Michihisa Koyama[2,3,4]

[1]Graduate School of Nanobioscience, Yokohama City University, 22-2 Seto, Kanazawa-ku, Yokohama 236-0027, Japan
[2]Graduate School of Engineering, Hiroshima University, 1-4-1 Kagamiyama, Higashi-Hiroshima 739-8527, Japan
[3]Research Initiative for Supra-Materials, Shinshu University, 4-17-1 Wakasato, Nagano, Nagano 380-8553, Japan
[4]INAMORI Frontier Research Center, Kyushu University, 744 Motooka, Fukuoka 819-0395, Japan

AUTHOR INFORMATION
**Corresponding Authors**
Michihisa Koyama: koyama_michihisa@shinshu-u.ac.jp
Takayoshi Ishimotol: tishimo@yokohama-cu.ac.jp

ABSTRACT

To understand the vibrational and configurational entropy effects for the stability of core-shell and solid-solution bimetallic nanoparticles, we theoretically investigated the excess energy of PdPt nanoparticles, adopting the $(PdPt)_{201}$ model of ca. 2 nm by using the density functional method. The vibrational energy and entropy terms contributed to the total energy of both core-shell and solid-solution nanoparticles. The configurational entropy term was defined only for the solid-solution nanoparticles. Although the absolute values of vibrational energy and entropy terms were much larger than that of configurational entropy term, their contributions were limited in the form of excess free energy due to the small difference between different atomic configurations. The large contribution of configurational entropy term to the excess free energy was clearly confirmed from our first-principles calculations. To estimate the stability of core-shell and solid-solution metal nanoparticles based on the excess energy, the configurational entropy term was the dominant factor.

## I. INTRODUCTION

Metal nanoparticles are widely used in environmental and energy related materials such as fuel cell electrocatalysts, automobile exhaust gas catalysts, and hydrogen storage materials [1-3]. The bimetallic alloy nanoparticles are often used to manipulate the physicochemical features by controlling the particle size, chemical composition, atomic configuration such as core-shell and solid solution [4-11].

Among a variety of combinations, PdPt bimetallic nanoparticles have been investigated intensively as unique catalyst [12-18]. Pd-Pt core-shell nanoparticles show higher catalytic activity than Pt monometallic nanoparticles [12-15]. Wang *et al.* pointed out that the weaker OH adsorption on Pt surface of Pd-Pt core-shell nanoparticle is one of the reasons [14]. They also mentioned that the stress of Pt surface due to mismatch between Pd and Pt interface is also an important factor to change the binding energy of OH. On the other hand, Pt-Pd core-shell nanoparticles are effective for oxidation reaction of formic acid, methanol and so on, compared with Pd or Pt monometallic nanoparticles [16-18]. Lei *et al.* pointed out that the large stabilization of Pt-Pd core-shell nanoparticle relates to the high catalytic activity [17].

In the above preceding studies, the stability of core-shell configuration is implicitly or explicitly assumed because of their immiscible nature from their bulk phase diagram. However, Kobayashi *et al.* found that Pd-Pt core-shell nanoparticle forms a homogeneously mixed solid-solution structure after the hydrogen adsorption/desorption process [19]. On the contrary, PdPt solid-solution was not formed from Pt-Pd core-shell structure even after the same hydrogen absorption/desorption process. The property of PdPt solid-solution was drastically changed. For example, the hydrogen storage capacity of PdPt solid-solution nanoparticle was larger than Pd monometallic nanoparticles, although the Pt, which does not absorb hydrogen under ambient temperature and pressure, is contained in the PdPt solid-solution nanoparticles. This indicates that unique properties, which are unexplored, may emerge, once combinations of immiscible metals are homogeneously mixed at the atomic level. Actually, various solid solution alloy nanoparticles of immiscible combinations have been synthesized, resulting in emergences of excellent properties [7,20-27]. Essential questions naturally arise; are the solid solution structures stable or semi-stable, does thermodynamics change in nanoparticle, what are key factors determining the miscibility and functionality?

In general, the thermodynamics of bulk alloy phase is determined by the enthalpy and entropy terms of both vibrational and mixing contributions. Recently, high-entropy alloys have attracted much attention [28-34]. Due to the large contribution of configurational entropy term with increasing the number of constituent elements, high-entropy alloys typically form homogeneous solid-solution structure at high temperature, while high-entropy alloys may form ordered alloy structure due to the decreased contribution from entropy term as the temperature decreases [34]. Likewise the bulk systems, the contribution of entropy is important in the thermodynamics of the nanoparticles. The stability of nanoparticles' atomic configurations, i.e., core-shell or solid-solution, is discussed based on the first-principles calculation [35,36]. Laasonen *et al.* investigated the stability of AgNi alloy of different atomic configurations with the size up to 1415 atoms while only the energy at 0 K is discussed [35]. Authors have investigated PdPt of different atomic configurations with the size of 711 atoms considering vibrational effect as well as configurational entropy [36]. However, the contribution of vibrational effect was not rigorously considered but was approximated by the values from classical molecular dynamics calculations of bimetallic systems [37], due to the high computational cost of frequency calculation within the framework of first-principles calculation. Even bimetallic nanoparticles, it is important to understand the contributions of vibrational and configurational entropy terms to the different atomic configuration systems.

In this study, we investigated the contributions of vibrational energy as well as the vibrational and configurational entropy terms for the stability of PdPt core-shell and solid-solution nanoparticles consisting of 201 atoms (ca. 2 nm) using the density functional theory (DFT) method.

## II. COMPUTATIONAL DETAILS

### A. DFT calculation

DFT calculations were performed using the Vienna ab initio simulation package (VASP) [38,39] with the projector augmented wave (PAW) method [40,41]. The Perdew-Burke-Ernzerhof (PBE) exchange and correlation functional was used under generalized gradient approximations [42]. The cutoff energy was set to be 400 eV. $1 \times 1 \times 1$ k-points were sampled by the Monkhorst-Pack grid method [43]. The PdPt core-shell and solid-solution nanoparticle structures were placed in a cubic cell for DFT calculation. The cell size was determined using periodic boundary conditions from a 10 Å vacuum region with particle size of model for all x, y, and z directions to avoid interactions between nanoparticles during the DFT calculation. To consider the vibrational energy and entropy of metal nanoparticle, the vibrational frequency for whole metal nanoparticle system is necessary. The displacement of 0.01 Å is used in the vibrational frequency calculation.

### B. Calculation of thermodynamic properties

Thermodynamics of metal nanoparticles is determined by the interplay between enthalpy and entropy:

$$G = H - TS = E_{el} + E_{vib} + RT - T(S_{vib} + S_{conf}), \qquad (1)$$

where $G$, $H$, and $S$ are Gibbs free energy, enthalpy, and entropy terms, respectively. $R$ and $T$ are gas constant and temperature, respectively. $E_{el}$ is obtained as the total energy by the first-principles calculation. $E_{vib}$ and $S_{vib}$ are vibrational energy and entropy terms, respectively. $S_{conf}$ is configurational entropy term. Here, translational and rotational terms of energy and entropy are neglected due to the small contribution compared with the vibrational terms. $E_{el}$ is obtained from the VASP calculation. $E_{vib}$ and $S_{vib}$ are calculated as

$$E_{vib} = RT \left\{ \sum_{i}^{3N-6} \frac{\theta_v^i}{T} \left( \frac{1}{2} + \frac{e^{-\frac{\theta_v^i}{T}}}{1 - e^{-\frac{\theta_v^i}{T}}} \right) \right\}, \qquad (2)$$

$$S_{vib} = R \sum_{i}^{3N-6} \left\{ \frac{\frac{\theta_v^i}{T} e^{-\frac{\theta_v^i}{T}}}{1 - e^{-\frac{\theta_v^i}{T}}} - \ln\left(1 - e^{-\frac{\theta_v^i}{T}}\right) \right\}, \tag{3}$$

where $\theta_v^i$ and $N$ are vibrational temperature and number of freedoms, respectively. The $\frac{\theta_v^i}{T}$ can be written as

$$\frac{\theta_v^i}{T} = \frac{h\omega_i c}{kT}, \tag{4}$$

where $h$, $c$, and $k$ are Planck's constant, speed of light, and Boltzmann's constant. The ω is vibrational frequency, which is calculated from DFT calculation. To obtain the vibrational frequency from DFT calculation, the $E_{vib}$ and $S_{vib}$ are calculated based on the theory.

The configurational entropy term by using Boltzmann's expression, $S_{conf}$, is defined as

$$S_{conf} = k \ln W, \tag{5}$$

where $W$ is the number of possible microscopic arrangements.

By using these equations, the temperature effect for the stability of metal nanoparticles are estimated in this study.

## C. Preparation of nanoparticle models

A truncated octahedral structure was adopted to represent the shape of the nanoparticles [44-46]. We used PdPt core-shell and solid-solution nanoparticles consisting of 201 atoms (ca. 2.0 nm) as calculation models. As core-shell and solid-solution nanoparticle models, four compositions were prepared: $Pd_{19}Pt_{182}$, $Pd_{79}Pt_{122}$, $Pd_{122}Pt_{79}$, and $Pd_{182}Pt_{19}$ (see Figure 1). In the case of a core-shell nanoparticle, Pd and Pt in $Pd_{19}Pt_{182}$ correspond to the core and the shell, respectively. $Pd_{19}Pt_{182}$ ($Pd_{182}Pt_{19}$) and $Pd_{79}Pt_{122}$ ($Pd_{122}Pt_{79}$) core-shell nanoparticles consist of 2-layer and 1-layer shell model, respectively. It was determined that these atomic compositions form clean core-shell structures. Based on these atomic components, solid-solution nanoparticles were prepared.

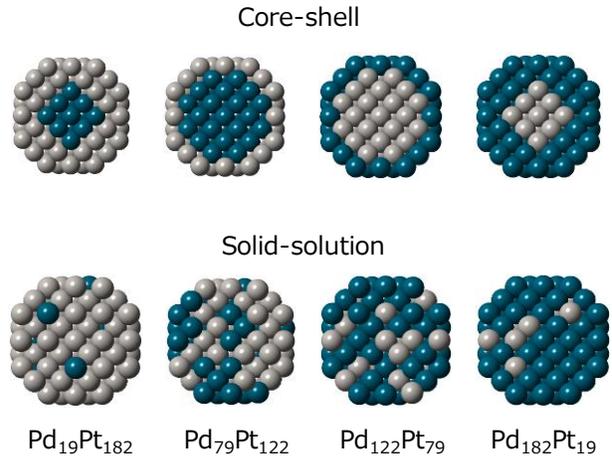

Fig. 1. Slice view of core-shell and solid-solution nanoparticles of $Pd_{19}Pt_{182}$, $Pd_{79}Pt_{122}$, $Pd_{122}Pt_{79}$, $Pd_{182}Pt_{19}$ after geometry optimization. Green and gray spheres are Pd and Pt atoms, respectively.

To evaluate the solid-solution nanoparticles, we introduced the Warren-Cowley (WC) short-range order parameter [47], which is a parameter to define the randomness of two-component systems. The WC parameter ($\alpha_i$) is defined by the following equation

$$\alpha_i = 1 - \frac{P_A^i}{C_A}, \tag{6}$$

where $C_A$ is the concentration of A atoms and $P_A^i$ is a conditional probability of having B atoms as neighbor in $i$-th coordination sphere. When $\alpha_i$ approaches zero, the A and B atoms are regarded to become complete randomness. The values of WC parameter for $Pd_{19}Pt_{182}$, $Pd_{79}Pt_{122}$, $Pd_{122}Pt_{79}$, and $Pd_{182}Pt_{19}$ were $8.94 \times 10^{-4}$, $1.49 \times 10^{-4}$, $3.31 \times 10^{-4}$, and $1.24 \times 10^{-4}$, respectively. The models prepared in this study were regarded as the solid-solution nanoparticles due to the small values of WC parameter.

## III. RESULTS AND DISCUSSION

### A. Excess energy of core-shell and solid-solution nanoparticles

After geometry optimization, the stability of core-shell and solid-solution nanoparticles was analyzed based on the electronic structure calculation. To discuss the stability of PdPt nanoparticles, we used the excess energy. The excess energy of each configuration, $E_{exe}$, is defined as

$$E_{exe} = \frac{1}{201}\left(E(\text{PdPt}) - \frac{N_{\text{Pd}}}{201}E(\text{Pd}) - \frac{N_{\text{Pt}}}{201}E(\text{Pt})\right), \tag{7}$$

where $N_{\text{Pd}}$ and $N_{\text{Pt}}$ are the numbers of Pd and Pt atoms in PdPt nanoparticles, respectively and they satisfy the sum rule, $N_{\text{Pd}} + N_{\text{Pt}} = 201$. $E(\text{PdPt})$ is the total energy of the core-shell or solid-solution nanoparticles. The $E(\text{Pd})$ and $E(\text{Pt})$ is the total energy of the $Pd_{201}$ and $Pt_{201}$ nanoparticles, respectively. When the excess energy is negative, the PdPt nanoparticle is energetically favored over the monometallic nanoparticles. Figure 2(a) shows the excess energy at 0 K of core-shell and solid-solution structures from monometallic Pd and Pt nanoparticles. All solid-solution structures were stable compared with the monometallic Pd and Pt nanoparticles. The Pd-Pt core-shell structures ($Pd_{19}Pt_{182}$ and $Pd_{79}Pt_{122}$) were unstab le compared with the corresponding solid-solution systems. On the other hand, Pt-Pd core-shell structures ($Pd_{122}Pt_{79}$ and $Pd_{182}Pt_{19}$) were more stable than the solid-solution nanoparticles. Comparing the surface energy of Pd and Pt, the surface energy of Pd(111) and Pd(100) is smaller than that of Pt(111) and Pt(100) [48]. This result indicates that the surface energy of PdPt nanoparticle tends to become smaller with increasing the number of Pd atoms on the surface at the same composition.

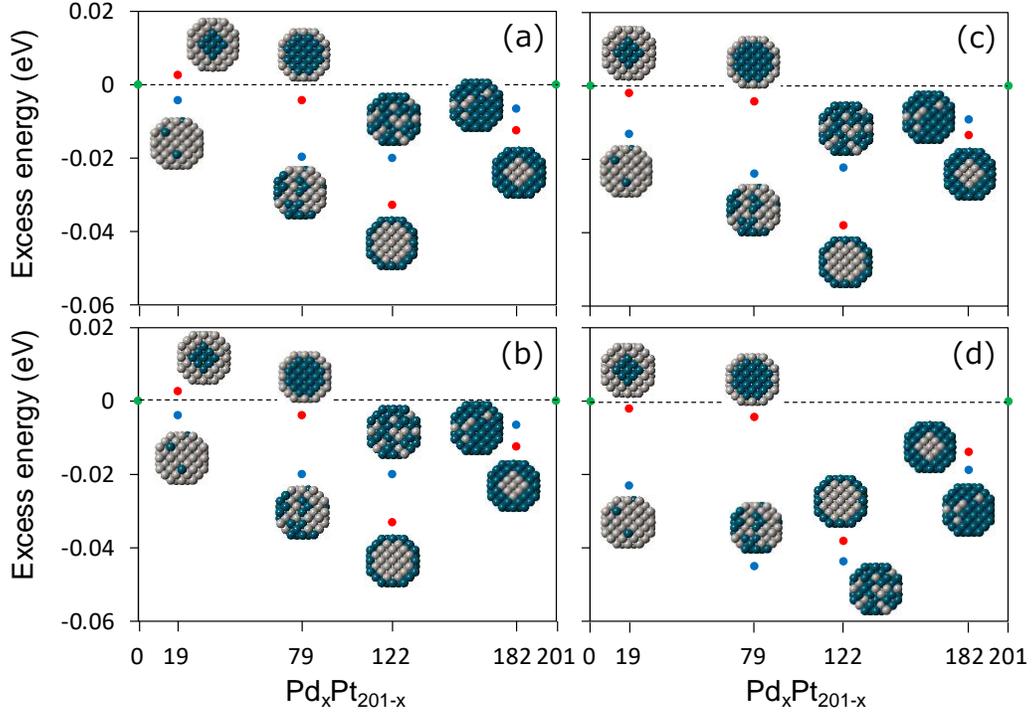

Fig. 2. Excess energy of core-shell (red circle) and solid-solution (blue circle) nanoparticles of $Pd_{19}Pt_{182}$, $Pd_{79}Pt_{122}$, $Pd_{122}Pt_{79}$, $Pd_{182}Pt_{19}$ from $Pd_{201}$ and $Pt_{201}$ nanoparticles. (a) Excess energy at 0 K, (b) excess energy with vibrational energy correction, (c) excess energy with vibrational energy and entropy correction, and (d) excess energy with vibrational energy and entropy and configurational entropy correction. Slice views of core-shell and solid-solution nanoparticles are shown. Green and gray spheres are Pd and Pt atoms, respectively.

## B. Vibrational energy and entropy and configurational entropy terms

The contribution of vibrational energy and entropy and configurational entropy terms is important to discuss the stability at the given temperature. We calculated those values from DFT calculation as shown in Table 1. Figure 3 shows the temperature dependence of vibrational energy, vibrational entropy, and configurational entropy terms of all nanoparticles investigated. Note that configurational entropy for monometal and core-shell nanoparticles is not shown because of the zero value. The largest value is the vibrational entropy term represented by triangles, followed by the vibrational energy represented by diamonds. Apparently, the configurational entropy term is the smallest. Focusing on the difference of each term between different configurations, the difference of vibrational energy is the smallest followed by the vibrational entropy term. Difference of configurational entropy is significant. To see more closely, we discuss the values at 373 K, which is a temperature of experimental condition [19]. The vibrational energy of nanoparticles was about $9.7 \times 10^{-2}$ eV. Vibrational energy did not change when number of Pd and Pt atoms or conformation of nanoparticle is different. The vibrational entropy term of nanoparticles was about $5 \times 10^{-4}$ eV/K. As well as the vibrational energy, vibrational energy also showed almost same value. The vibrational entropy term at 373 K was larger than vibrational energy. This result indicates that the vibrational entropy term largely affects the stability of nanoparticle itself. On the other hand, the configurational entropy term is meaningful only for solid solution structures. Contrary to the vibrational energy and entropy terms, configurational entropy term was strongly dependent on the number of Pd and Pt atoms and conformation. While the configurational entropy term was zero for core-shell structures, the solid-solution structures shows a certain value of the configurational entropy term. In addition, when the Pd-Pt compositions is close to equal, the configurational entropy term becomes larger. For example, the configurational entropy term of $Pd_{79}Pt_{122}$ solid solution structure was larger than that of $Pd_{19}Pt_{182}$. The largest contribution to the total energy was the vibrational entropy term at 373 K. The vibrational energy was about half of the vibrational entropy term. The configurational entropy term of solid-solution structures was less than one-tenth compared with the vibrational entropy.

Table 1. Vibrational energy ($E_{vib}$) and entropy ($S_{vib}$) and configurational entropy ($S_{conf}$) terms of core-shell (CS) and solid-solution (SS) nanoparticles of $Pd_{19}Pt_{182}$, $Pd_{79}Pt_{122}$, $Pd_{122}Pt_{79}$, $Pd_{182}Pt_{19}$ at 373 K. The values of $E_{vib}$, $S_{vib}$, and $S_{conf}$ are shown per atom.

| Pd | Pt | geometry | $E_{vib}$ (eV) | $S_{vib}$ (eV/K) | $S_{conf}$ (eV/K) | $-TS_{vib}$ (eV) | $-TS_{conf}$ (eV) |
|---|---|---|---|---|---|---|---|
| 0 | 201 | | $9.74 \times 10^{-2}$ | $5.01 \times 10^{-4}$ | 0 | -0.187 | 0.000 |
| 19 | 182 | CS | $9.72 \times 10^{-2}$ | $5.13 \times 10^{-4}$ | 0 | -0.192 | 0 |
| 19 | 182 | SS | $9.71 \times 10^{-2}$ | $5.15 \times 10^{-4}$ | $2.59 \times 10^{-5}$ | -0.192 | -0.010 |
| 79 | 122 | CS | $9.74 \times 10^{-2}$ | $5.01 \times 10^{-4}$ | 0 | -0.187 | 0 |
| 79 | 122 | SS | $9.72 \times 10^{-2}$ | $5.11 \times 10^{-4}$ | $5.67 \times 10^{-5}$ | -0.191 | -0.021 |
| 122 | 79 | CS | $9.73 \times 10^{-2}$ | $5.13 \times 10^{-4}$ | 0 | -0.192 | 0 |
| 122 | 79 | SS | $9.73 \times 10^{-2}$ | $5.06 \times 10^{-4}$ | $5.67 \times 10^{-5}$ | -0.189 | -0.021 |
| 182 | 19 | CS | $9.74 \times 10^{-2}$ | $5.03 \times 10^{-4}$ | 0 | -0.188 | 0 |
| 182 | 19 | SS | $9.74 \times 10^{-2}$ | $5.07 \times 10^{-4}$ | $2.59 \times 10^{-5}$ | -0.189 | -0.010 |
| 201 | 0 | | $9.74 \times 10^{-2}$ | $5.00 \times 10^{-4}$ | 0 | -0.187 | 0 |

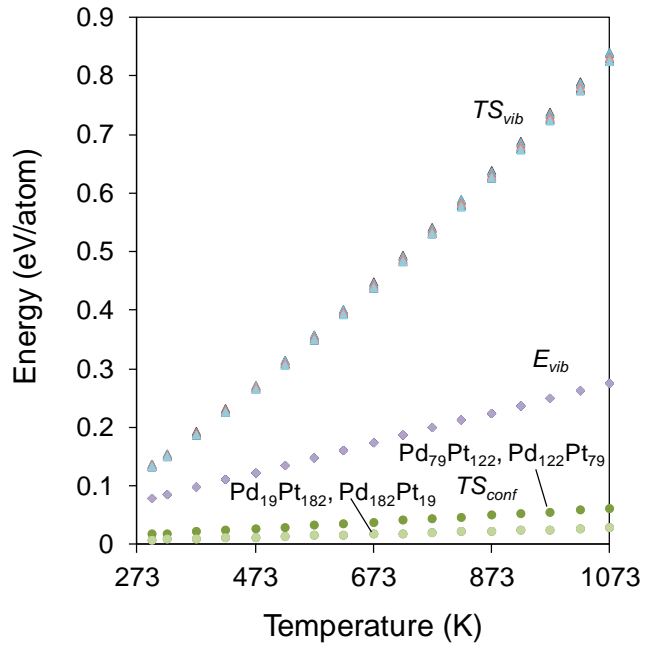

Fig. 3. Temperature dependence of vibrational energy (diamond), vibrational entropy (triangle), and configurational entropy (circle) terms. Those for all configurations are shown except for configurational term for monometal and core-shell nanoparticles.

## C. Contribution of temperature effect to excess energy

We analyzed the contributions of temperature effect, such as vibrational energy, vibrational entropy, and configurational entropy, to the excess energy of sore-shell and solid-solution structures. Table 2 shows the calculated values of excess energy and temperature effect of core-shell and solid-solution structures. The excess energy in this table means the values at 0K shown in Fig. 2(a). The values after temperature correction are calculated values for each term. The contribution of vibrational energy to the excess energy was almost zero although the absolute value of vibrational energy was about $9.7 \times 10^{-2}$ eV. Because the values of vibrational energy of core-shell, solid-solution and monometallic structures were almost the same, the contribution of vibrational energy to the excess energy was cancelled each other. Compared to the vibrational energy, the contribution of vibrational entropy term of core-shell, solid-solution and monometallic structures was larger. In case of configurational entropy term, the contribution of configurational entropy term of solid-solution was much larger than other terms, because the configurational entropy term of reference monometallic structures was zero. Figure 4 shows temperature dependent terms in excess free energy of solid-solution and monometal nanoparticles. As discussed above, absolute values of vibrational enthalpy and entropy terms are large while the difference between monometal and solid-solution nanoparticle results in the limited contribution to the stabilization. On the contrary, configurational entropy term shows the largest contribution to the stabilization at any temperature shown in Fig. 4. When we focus on the equal composition region, one can clearly see that the stabilization of solid structure nanoparticles is mostly contributed by the configurational entropy contribution.

Table 2. Excess energy of core-shell (CS) and solid-solution (SS) nanoparticles of $Pd_{19}Pt_{182}$, $Pd_{79}Pt_{122}$, $Pd_{122}Pt_{79}$, $Pd_{182}Pt_{19}$ and contribution of vibrational energy ($E_{vib}$) and entropy ($S_{vib}$) and configurational entropy ($S_{conf}$) terms at 373 K. The values of $E_{vib}$, $S_{vib}$, and $S_{conf}$ are shown per atom.

| Pd | Pt | geometry | $E_{exe}$ (eV) | $E_{vib}$ (eV) | $-TS_{vib}$ (eV) | $-TS_{conf}$ (eV) |
|---|---|---|---|---|---|---|
| 0 | 201 |  | 0 | 0 | 0 | 0 |
| 19 | 182 | CS | $0.27 \times 10^{-2}$ | $-2.48 \times 10^{-4}$ | $-4.46 \times 10^{-3}$ | 0 |
| 19 | 182 | SS | $-0.77 \times 10^{-2}$ | $-2.87 \times 10^{-4}$ | $-5.08 \times 10^{-3}$ | $-0.97 \times 10^{-2}$ |
| 79 | 122 | CS | $-0.41 \times 10^{-2}$ | $0.19 \times 10^{-4}$ | $-0.14 \times 10^{-3}$ | 0 |
| 79 | 122 | SS | $-1.97 \times 10^{-2}$ | $-1.56 \times 10^{-4}$ | $-3.88 \times 10^{-3}$ | $-2.11 \times 10^{-2}$ |
| 122 | 79 | CS | $-3.29 \times 10^{-2}$ | $-2.35 \times 10^{-4}$ | $-4.73 \times 10^{-3}$ | 0 |
| 122 | 79 | SS | $-2.01 \times 10^{-2}$ | $-0.77 \times 10^{-4}$ | $-2.09 \times 10^{-3}$ | $-2.11 \times 10^{-2}$ |
| 182 | 19 | CS | $-1.25 \times 10^{-2}$ | $0.35 \times 10^{-4}$ | $-1.06 \times 10^{-3}$ | 0 |
| 182 | 19 | SS | $-0.65 \times 10^{-2}$ | $0.49 \times 10^{-4}$ | $-2.44 \times 10^{-3}$ | $-0.97 \times 10^{-2}$ |
| 201 | 0 |  | 0 | 0 | 0 | 0 |

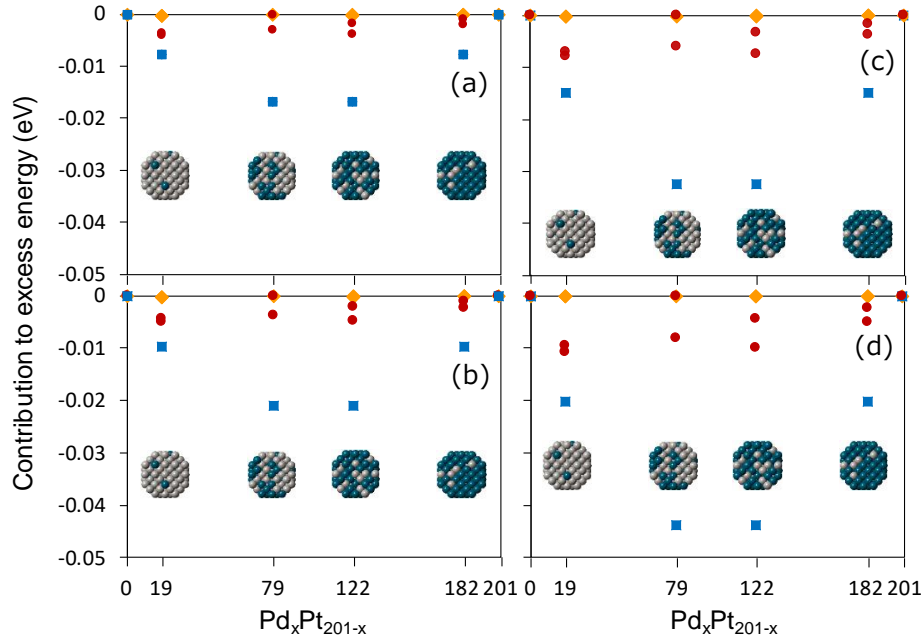

Fig. 4. Contribution of vibrational energy (yellow diamond), vibrational entropy (red circle), and configurational entropy (blue square) terms to excess energy at selected temperatures of (a) 298K, (b) 373K, (c) 573K, and (d) 773K.

Based on the temperature correction of core-shell and solid-solution nanoparticles, we plotted the excess energy in Figs. 2(b)-(d). The excess energy with vibrational energy correction (see Fig. 2(b)) was the same as the excess energy without correction (Fig. 2(a)). By adding the vibrational energy and entropy corrections to the excess energy (Fig. 2(c)), some structures were slightly more stabilized. For example, $Pd_{19}Pt_{192}$ core-shell structure showed negative excess energy. The $Pd_{19}Pt_{182}$ solid-solution structure also stabilized. However, overall trend of excess energy of Pd-Pt core-shell and solid-solution structures were almost the same. Further addition of configurational entropy term to the excess energy (Fig. 2(d)), we clearly obtained drastic change of the stability of core-shell and solid solution structures. Due to the large contribution of configurational entropy term, the solid-solution structures became more stable. The excess energy difference between solid-solution and Pd-Pt core-shell structures became larger. The solid-solution structures were more stable than the Pt-Pd core-shell structures, although the Pt-Pd core-shell structures were more stable compared with the PdPt solid-solution without configurational entropy term. This result indicates that the configurational entropy term plays an critical role in estimating the stability of core-shell and solid-solution nanoparticles at a given temperature. As seen from Fig. 4, the contribution of configurational entropy term becomes larger with increasing temperature.

## IV. CONCLUSIONS

In this study, we theoretically investigated the stability of core-shell and solid-solution nanoparticles, using the $(PdPt)_{201}$ model, by using the DFT approach. The stability of PdPt core-shell and solid-solution nanoparticles was analyzed by the excess energy. The stability of PdPt core-shell and solid-solution nanoparticles were affected by the difference of surface energy of monometallic Pd and Pt nanoparticles. The vibrational energy and entropy terms were contributed to the total energy of metal nanoparticles, although the number of Pd and Pt atoms and configuration did not affect the vibrational energy and entropy terms significantly. On the contrary, the configurational entropy term of the solid-solution structures depends on the composition. The contributions of vibrational energy and entropy terms to excess energy were limited because the similar values of alloy and reference monometals cancel. On the contrary, a large contribution of configurational entropy term to the excess energy was noticed. It was clarified that the configurational entropy term was the dominant factor for the stability of core-shell and solid-solution metal nanoparticles at a finite temperature.

**ACKNOWLEDGMENTS**

This work was supported by JST-CREST, and JST-ACCEL (JPMJAC1501), and the Advanced Computational Scientific Program of the Research Institute for Information Technology, Kyushu University.